\documentstyle[twocolumn,aps]{revtex}
\input epsf
\begin{document}
\draft
\twocolumn[\hsize\textwidth\columnwidth\hsize\csname@twocolumnfalse\endcsname

\title{Dynamics of the Number of Trades of Financial Securities}
\author{Giovanni Bonanno, Fabrizio Lillo and Rosario N. Mantegna}
\address{
Istituto Nazionale per la Fisica della Materia, Unit\`a di Palermo\\
and\\ Dipartimento di Energetica ed Applicazioni di Fisica,Universit\`a di Palermo, 
Viale delle Scienze, I-90128, Palermo, Italia}
\maketitle
\begin{abstract}
We perform a parallel analysis of the spectral density of (i) the logarithm of
price and (ii) the daily number of trades of a set of stocks traded in the New
York Stock Exchange. The stocks are selected to be representative of a wide
range of stock capitalization. The observed spectral densities show a different
power-law behavior. We confirm the $1/f^2$ behavior for the spectral density of
the logarithm of stock price whereas we detect a $1/f$-like behavior for the
spectral density of the daily number of trades.
\end{abstract}
\vskip2pc]

\tighten
\narrowtext

\section{Introduction}

In the field of econophysics \cite{Ms99} several empirical researches 
have been performed to investigate the statistical properties of stock
price  and volatility time series of assets traded in financial markets
(for a recent overview see for example \cite{Mantegna99,Bouchaud2000}).
Comparably less attention has been devoted to the investigation of
statistical properties of the dynamics of the number of trades of a 
given asset. A similar investigation is relevant for the basic
problem of a quantitative assessment of the liquidity of the 
considered asset. There are two general aspects of liquidity for an
asset traded in a market. The first concerns how easily assets can 
be converted back into cash in a market, whereas the second refers
to the general problem of the ability of the market to absorb a 
given amount of buying and selling of shares of a given security 
at reasonable price changes. In the present study we consider the 
statistical properties of trading 
related to the second aspect of liquidity cited above. Specifically 
we investigate the temporal dynamics of the number of daily trades of 
88 stocks traded in the New York Stock Exchange. The 88 stocks are selected at 
different values of their capitalization to provide a set of stocks 
representative of different levels of capitalizations. The interval
of capitalization of our set is spanning more than three orders of magnitude 
in capitalization. With this choice  we are able to investigate both
heavily traded and less frequently traded stocks. In the present
study we focus our attention on the time memory of the dynamics of the
number of trades of a set of financial securities. We find that 
most capitalized stocks are characterized by a dynamics of the number
of trades having interesting statistical properties. The time evolution
of the number of trades has associated a power spectrum which is 
well approximated by a $1/f$-like behavior. We interpret this result
as a quantitative evidence that the level of activity of highly 
capitalized and frequently
traded stocks in a financial market does not possess a typical time 
scale. The same behavior is also qualitatively detected in less
capitalized stocks although the value of the exponent $\gamma$ of the
power spectrum $S(f) \propto 1/f^{\gamma}$ deviates from the value 
$\gamma=1$ and moves 
to lower values for stocks with lower capitalization. The modeling 
of systems with $1/f$ power spectra is still a theoretical 
unsolved problem. Hence the level 
of activity of a given asset in a financial market presents a 
statistical behavior whose theoretical modeling 
is certainly challenging.
The paper is organized as follows: in the next section we illustrate
the analyses performed by discussing the results obtained for one of the
most capitalized stocks of the New York Stock Exchange. Section 3
presents the results obtained by investigating the selected set of
stocks and in Section 4 we briefly state our conclusions.

\section{Single stock analysis}

\indent We analyze the number of daily trades for a selected set
of securities traded in the New York Stock Exchange (NYSE).
Our database is the trade and quote (TAQ) database \cite{taq}
for the three-year period from Jan. 1995 to Dec. 1997. 
The typical number of daily trades $n(t)$ depends on the stock considered and
varies from a few to few thousands units per day.
We first analyze the dynamics of the number of daily trades for the
General Electrics Co. (GE), which is one of the most capitalized stocks in the NYSE 
in the investigated period. The time series of the daily number
 of trades is
shown in the inset of Fig. 1. The number of daily trades fluctuates
and the time series is non-stationary. In order to test the presence of long range correlation in this time
series, we determine its spectral density. The spectral density for
the time series of the number of trades of the GE is shown in the upper 
curve of Fig. 1. For comparison we also show the spectral density of
the logarithm of the daily price of the same security.  
We fit both spectral densities with a power-law function 
$S(f)\propto 1/f^{\gamma}$ using a logarithmic
binnig in order to not overestimate the high-frequency components. Our best 
estimation for the exponent $\gamma$ of the spectral density of the logarithm 
of the price is 
$\gamma = 1.93$. 
\begin{figure}[t]
\epsfxsize=3in
\epsfbox{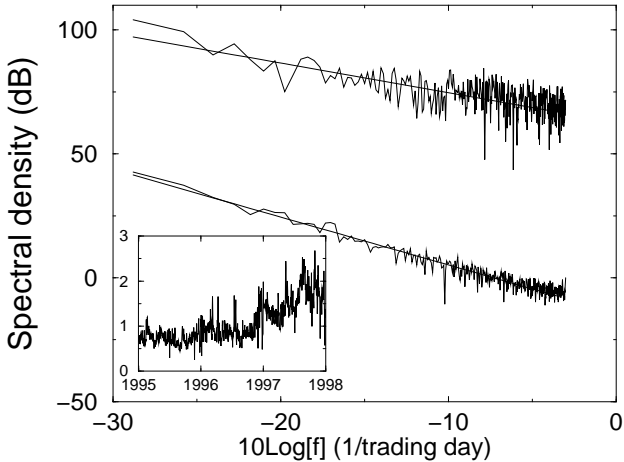}
\caption{Log-log plot of the spectral density  $S(f)
$ of the logarithm of price (bottom curve) and of the number
of daily trades (top curve) of the General Electrics Co. (GE). 
The two spectral densities are well approximated by a power-law behavior
$S(f)\propto 1/f^{\gamma}$ with $\gamma=1.93$ and $\gamma=1.19$, respectively.
The two straight lines are our best fit performed as explained in the text.
In the inset we show the time evolution of the number
of daily trades of GE (in thousands of trades) in the three-year 
investigated period.}
\label{fig1}
 \end{figure}
Hence, the spectral density is well approximated 
by $S(f)\sim 1/f^2$, which is the prediction for the spectral 
density of a random walk. This form of the spectral density is related 
to the fact that returns of the price are short time correlated. 
This result is well-established in financial literature
\cite{Cootner,Samuelson}. 
We observe a different behavior for the spectral
density of the daily number of trades. Our best fit for the 
exponent $\gamma$ gives $\gamma = 1.19$. This
value of the exponent is compatible with a $1/f$ noise \cite{Press,Keshner}.
The time series of the GE daily trades is non stationary and shows a 
clear trend. 
We check the effect of these features on the fitting of the spectral density 
by investigating surrogate data obtained removing the trend 
by evaluating a running average of the original time series 
computed using a window of length $L$. With this procedure, we 
obtain a detrended time series by dividing the original number 
of trades at a given day $t$ by the average of the number of 
trades in a period of length $L$ centered in $t$. We
take the values of $L$ equal to $11, 21, 41$ and $81$ trading days. 
The new time series do not present a global trend. In Fig. 2 we 
scomponents lower than $ 1/L$ are affected by the running average 
procedure. We perform a power law fit in the interval of
frequency higher than $1/L$ and our best fit for the exponent $\gamma$ gives
$\gamma= 1.09, 0.85, 0.85 ,0.91$ for $L=11,21,41,81$, respectively. 
The estimation of the exponent is weakly affected by the running 
averaging procedure. These results for the spectral density in 
the original and detrended time series indicate that the number 
of trades is intrinsically long-range correlated.

\begin{figure}[t]
\epsfxsize=3in
\epsfbox{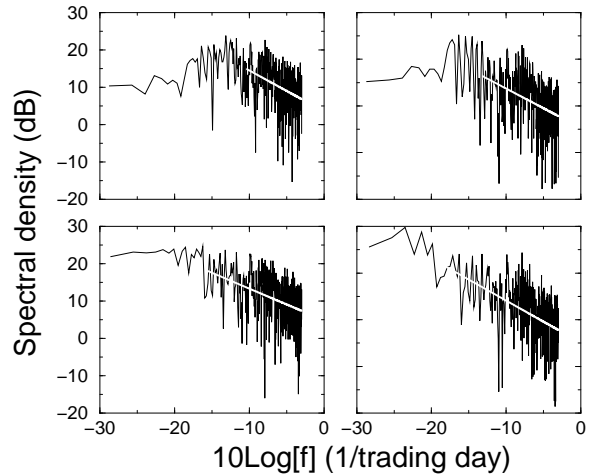}
\caption{Spectral densities of the number of daily trades obtained by
detrending the original time series. The detrending is obtained by dividing the
time series by the running average in a time window of length $L$. In the 
four panels we show the spectral densities obtained by detrending the original
time series with averaging windows of length 
$L=11$ (top left), $L=21$ (top right), $L=41$ (bottom left) and $L=81$
(bottom right) trading days. In the regions not affected by the detrending
procedure ($f>1/L$) a $1/f$-like behavior is always observed (solid line).}
\label{fig2}
\end{figure}

\section{Comparison between stocks}

In this section we study the general validity of the results 
illustrated in the previous section by studying a specific stock. 
In particular we determine the 
spectral density of the logarithm of the price and of the 
number of daily trades for a set of $88$ selected
stocks traded in the NYSE. We select the stocks randomly 
in a wide range of capitalization. The capitalization of our 
set ranges from $32\cdot 10^6$ USD, which is the capitalization of
the Sun Energy (SLP), to $80\cdot10^9$ USD for the GE.   
We determine the spectral density for the $\ln S_i(t)$ 
and of the $n_i(t)$, 
where $S_i(t)$ and $n_i(t)$ are the closure price and the
number of trades of company $i$ at day $t$, respectively. The subscript $i=1,...,88$ labels 
the company. 
For each spectral density we perform a power-law fit $S(f)\propto
1/f^{\gamma}$. 
We check the robustness of our fittings by determining
their correlation 
coefficient and we find absolute values ranging from $0.87$ to $0.99$ 
(average value $0.97$) for the 
logarithm of the price and from
$0.56$ to $0.96$ (average value $0.86$) for the number of trades.
In Fig. 3 we show the two exponents as a function of the
capitalization of the investigated security. The large majority
of values of the exponent $\gamma$ of
the spectral densities of the logarithm of the price is close to the value
$\gamma=2$. On the other hand the large majority
of values of the exponent $\gamma$ of the spectral
densities of the number of daily trades is close to the value $\gamma=1$, in
particular for the most capitalized stocks.
Moreover, the exponent $\gamma$ increases towards the value $\gamma=1$ 
as far as the capitalization increases.
When we consider less capitalized stocks the typical number of daily 
trades can be very small, of the order of $10$ or less. This fact 
leads to a problem of high digitalization noise of the time series, 
\begin{figure}[t]
\epsfxsize=3in
\epsfbox{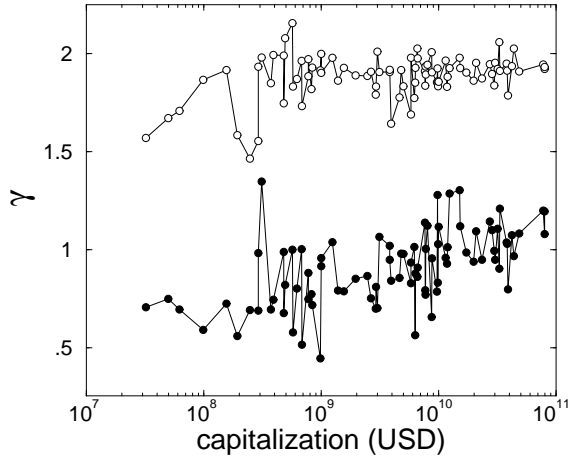}
\caption{Plot of the exponent $\gamma$ of the power-law fit $S(f)\propto
1/f^{\gamma}$ of the spectral density of the number of trades (filled circle)
and of the logarithm of the price (empty circle) as a function of the
capitalization of the investigated company. The exponent $\gamma$ for
the logarithm of the price is almost constant for the majority of stocks and
clusters in the neighborhood of $\gamma=2$. On the contrary,
for the number of trades, the exponent $\gamma$ slowly increases
by increasing the capitalization and clusters in the neighborhood 
of $\gamma=1$ for the most capitalized stocks.}
\label{fig3}
 \end{figure}
because the number of trades is an integer
and the quantization noise is relevant. As a consequence the 
statistical analysis on these
time series may present a non negligible white noise contribution. 
This fact may be one reason for the observed decreases of the 
value of the exponent $\gamma$ in less capitalized stocks.

\section{Conclusion}

The statistical properties of (i) the logarithm of the price and (ii) the number
of daily trades of a security traded in a financial market are quite different.
Specifically, we confirm that the time series of the logarithm of the price are
characterized by a $1/f^2$ spectral density, whereas we observe that the time
series of the daily number of trades show a $1/f$-like spectral density. The 
$1/f^2$ behavior observed in the time evolution of the logarithm of the price
reflects the short time temporal memory of price returns. This kind of
short-range time memory is needed to ensure absence of arbitrage opportunities
in the market.
On the other hand the $1/f$-like behavior observed in the daily number of trades
manifests the absence of a typical scale in the time memory of this variable. In
other words the activity of trading is not constant in time even for most
capitalized stocks and its modeling needs to take into account phenomena
occurring at a variety of time scales. Realistic models of trading activities in
financial markets should take into account this feature empirically observed.

\section{Acknowledgements}

The authors thank INFM and MURST for financial support. This work 
is part of the FRA-INFM project 'Volatility in financial markets'. 
G. Bonanno and F. Lillo acknowledge FSE-INFM for their fellowships.


\end{document}